# Ocean Terracing

**Richard B. Cathcart**
Geographos
1300 West Olive Avenue, Suite M
Burbank, California 91506, USA
rbcathcart@charter.net

**Alexander A. Bolonkin**
C & R
1310 Avenue R, Suite 6-F
Brooklyn, NY 11229, USA
aBolonkin@juno.com

**ABSTRACT**

Artworks can improve humanity's ability to apply macro-engineering principles which skirt or correct oceanographic problems impairing the economic usefulness of coastal land, the overhead airshed, and seawater's temperature and salinity stability. A new form of Art, "Ocean Art", is here proposed which centers on deliberate terracing of appropriate regions of our world's ocean; a proposed example of macro-engineered useful "Ocean Art" is the technically possible 21$^{st}$ Century terracing of the Mediterranean Sea. "Ocean Art" is applicable worldwide to places that might be practically improved by its judicious employment. Such "Ocean Art" may constitute an entirely unique category of solutions to coastal disaster prevention planning.

## 1. INTRODUCTION

Space Art proponents opt to construct various symbolic artifacts in Outer Space visible from Earth's surface (Bureaud, 2005). With plastic film and textile envelopes, Air Art advocates exploit the possibilities of compressed air or naturally generated atmospheric wind (Cathcart and Cirkovic, 2006). Land Art results from different human interpretations of the significance of natural and anthropogenic sub-aerial landscapes (Tiberghien, 1995). As Constanze Guthenke (2006) stresses: "Grammatically…the 'Mediterranean' denotes a quality, rather than an object, an indicator of place, rather than the place itself" because the OXFORD ENGLISH DICTIONARY's editors consider the word primarily an adjective.

Famed artist Christo installed 11 flamingo-pink floating plastic collar-mats covering approximately 600,000 m$^2$ of a saltwater lagoon surface in Miami, Florida's Biscayne Bay for his temporary "Surrounded Islands Project" artwork of 1983. During 1969, Peter Hutchinson and Dennis Oppenheim installed artworks in the coastal waters off Tobago in the West Indies. "Wave Organ", constructed by Peter Richards and George Gonzales, and the seashore artworks of Andy Goldsworthy are also examples of what is, charitably, labeled Oceanographic Art since they are essentially decorative. However, Ocean Art is a form of seawater sculpting by aquatic terracing focused on the 70% of Earth's surface that is oceanic; the modern-day originator of geographically large-scale intentional Ocean Art is the German architect Frei Otto, who first contemplated the concept in 1953 (Nerdinger, 2005). Ocean Art has a practical, commercially useful aspect that Oceanographic Art does not and, therefore, is of interest to 21$^{st}$ Century adherents of Ocean Macro-engineering.

## 2. INADVERTANT OCEAN TERRACING

About 5,000 years ago, the Earth-atmosphere's methane concentration began to increase. Its generation source was human cultivation of rice in flatland paddies; about 2,000 years ago, humans commenced growing rice in watery paddies on terraced hillsides. Methane is a greenhouse gas and the anthropogenic contribution to the Earth-atmosphere causes global warming and a rise worldwide of sea level (Ruddiman, 2005). Perhaps half of all living humans eat rice and the microorganisms living in anoxic rice field soils contribute between 10% and 25% of annual global methane emissions (Lu, 2005); by 2030, there might be five billion human consumers of rice (Khush, 2005). Artificial wetlands emplaced on shaped hillsides have in the past, and continue in the present, to contribute to the global ocean's sea level instability (Cheng, 2006). Catastrophic methane releases, caused by submarine slumping during which submarine sliding of bottom material results in hydrate dissociation triggered by depressurization (sea level fall), have occurred in the Mediterranean Sea Basin (Maslin et al. 2004). An anthropogenically-induced maintenance of the Mediterranean Sea's level at one meter below the prevailing future World-Ocean level is unlikely, even with a possible contributory permanent seawater-warming steric phenomenon (Garcia, 2006), to trigger a large methane release from the seafloor during the $21^{st}$ Century.

Especially subsequent to the Industrial Revolution, dams on rivers were constructed that have the effect of retaining freshwater in large volumes. There are probably 45,000 dams exceeding 15 m high that, in the aggregate, can store more than 6,500 $km^3$ of freshwater (Nilsson, 2005). (The Netherlands' famous Ijsselmeer did not directly terrace the ocean, but simply replaced a segment of seawater with an equal volume of freshwater; France's La Rance Tidal Barrage merely harnesses the daily tides.) Absent these on-land reservoirs, the ocean would be higher than today. Anthropogenic global warming of the atmosphere caused, in part, by a measured build-up of carbon dioxide gas during the past two centuries, is plausibly suggested to also cause a global sea level rise, which will initially be regionally differentiated (Raper and Braithwaite, 2006; Khilyuk and Chilingar, 2006). Evidently, and prospectively, it is *Homo sapiens'* intention to occupy the land and the ocean with various infrastructures; *per contra*, humans can still only use the Earth's air.

## 3. OCEAN TERRACING FACTORS

By 2100, the world's ocean could rise by 0.5 - 1.0 m relative to its present-day level thereby directly affecting the world's coastline (Crossland, 2005). Humanity's activities (to make and earn a living) will "globalize" the Mediterranean Sea—its seawater, organic and inorganic contents, and coast. All coastal nations should now be planning for 0.5 - 1.0 m rise in sea level during the $21^{st}$ Century. Nations of the Mediterranean Sea Basin have several old and expensive ameliorative macroprojects options available (Cathcart, 2006). However, recent R&D and newer products derived from advanced material technologies—particularly, technical textiles exhibiting high-performance, purely functional, and precisely woven or non-woven fabrics—offers the prospect of cheap regional sea level rise control macroprojects (Cullen, 2005; Brownell, 2006).

Commercial shipping interests and mass tourism organizers and related industries (Mosedale, 2004) are the most obvious pro-amelioration macroproject constituencies examining the impacts of future sea level rise in the Basin in addition to the ecologically sustainable fisheries (Goren and Galil, 2005), recreational boating and yachting, harbor maintenance and the military communities.

The Mediterranean Sea Basin, renowned for its beautiful beaches and healthful climate, exhibits boundary layer air pollution originating mostly from other regions of the world because the Basin lies at the "crossroads" of global wind currents and receives an inordinate amount of industrial pollutants (Lelieveld et al., 2002). Ship emissions contribute substantially to air pollution over the summertime Basin, which also modifies the region's cloud properties affecting the region's climate radiation budget through sulfate aerosol forcing (Marmer and Langmann, 2005). Nowadays, seawater in the Mediterranean Sea has visibly and measurably different characteristics than in the recorded past. Remarkable shifts in the thermohaline circulation and water mass characteristics have occurred because of mankind's activities on land, including large freshwater reservoir construction (Skliris and Lascaratos, 2004) and the widespread regulation of river sediment deposition (Liquete et al., 2004).

Peak evaporation from the Mediterranean Sea, driven by energy released from seawater, takes places during the wintertime (Fenoglio-Marc, 2006). The Mediterranean Sea measurably warmed during the 20$^{th}$ Century; the causes and potential effects of the elevated seawater temperature on pelagic food webs and the Basin's present-day and future climate are not well known. The Earth-atmosphere's response to an idealized 2 K cooling of the Mediterranean Sea has been estimated, but not the reverse, a warming (Li, 2006). Unmonitored Basin seafloor volcanoes, which have in the past created new land for a short period (Graham Island in 1831) and may create new land in the future (possibly at the Marsili seamount north of Sicily), are indicative of significant seafloor geothermal heating (Mullarney. 2006). Standard scientific explanations for imposed seawater warming in the Basin rely mostly on the alleged enhanced atmospheric global warming theory. During the 21$^{st}$ Century, Mediterranean Sea Basin hurricanes possibly may become slightly more intense owing to seawater's steric warming, with higher boundary layer wind speeds, and increased vertical mixing of seawater masses. A Basin-wide sea level higher by 1 m may also lead to higher wintertime storm surges affecting the Basin's coast and its costly-to-repair/replace infrastructure (Emanuel, 2005).

## 4. BRIDGE AND BARRAGE MACROPROJECTS

The Strait of Gibraltar connects the North Atlantic Ocean and the Mediterranean Sea. Long-span bridges are becoming longer because the use of supercomputers permits macro-engineers to usefully calculate the physical forces impinging—such as flutter stability (Astiz, 1998)—these long-span structures with greater accuracy and because of modern construction techniques that can isolate suspension bridge tower bases from earthquakes. Macro-engineers mostly must meet managerial challenges such as financing and resource accumulation. Technical challenges remain, of course, but new materials—such as carbon fibers embedded in composite materials (Meier, 1987; Noistering, 2000)—are being developed as practical solutions to such impending macro-problems.

Three-dimensional, supercomputer-generated bathymetric maps and high-resolution geologic sections (based on sparker profiles and sea-bottom drill cores) are already available that generally illustrate the Strait of Gibraltar's geologic framework today and during past periods of Geological Time. Anticipation of a Punta Malabata, Morocco to Punta Paloma, Spain Gibraltar Strait Tunnel, with construction tentatively slated to commence in 2008, has provoked all macroproject site investigation efforts so far. Additional pre-planning site studies may be required if our proposed Gibraltar Strait Textile Barrage (GSTB) textile/plastic macroproject proposal is seriously considered. GSTB will be judiciously draped on a general alignment between Tarifa (36$^0$ 01' North by 5$^0$ 36' West) in Spain and Ksar e' Sghir (35$^0$ 50' North by 5$^0$ 32' West) in Morocco (Cerda, 2004), creating an aerial and submarine artwork somewhat imitative of Christo's "Valley Curtain, Rifle, Colorado, 1970-1972" (Vaizey, 1990). GOTO: **http://www.christojeanneclaude.net/vc.html**; it will bow or "billow" like a ship's sail eastwards from the selected 20 km alignment because of marine (difference in sea level on a two-sided bottom-anchored membrane and natural currents such as tidal solitons coming from the North Atlantic Ocean) and aerial (seasonal winds) pressures acting directly upon the GSTB. [Ernest C. Harris (1915-98) macro-engineered "Valley Curtain".] Planners of the GSTB macroproject will draw on the installation experience with heavy wire nets, floatation systems and their moorings derived from World War II anti-submarine net installations in strategic harbors and that documented experience offered by the 100 km-long World War I anti-submarine Otranto Strait Barrage (1915-19). One of the main factors influencing the GSTB's cost will be the sea-bottom cut-off wall to minimize seawater seepage—there is some slight possibility a submarine cut-off trench need not be dug on the seafloor, nor an uninterrupted underwater grout curtain need be installed to ensure proper functioning of the completed Gibraltar Strait Textile Barrage.

From its western approaches, the Gibraltar Strait Textile Barrage will have the characteristic of an architectural deception resembling an English Garden or zoo landscape architect's geotextile ha-ha (also known as a "sunken fence") in that—absent warning light-buoys and radar reflectors—ship navigators will visually misapprehend the true nature of the sea route ahead. Those mariners (such as private-sector fishermen and yachtsmen) piloting their boats and other types of watercraft without benefit of up-to-date navigational sea charts that indicate the GSTB's presence in the seascape will have no inkling via normal optical clues whatsoever that a 1 m drop in sea level obstructs the Strait of Gibraltar! Mariners without radar readouts using the eastern approaches will visually spy a 1 m-high tensioned fabric wall, which if made of clear or aquamarine-colored material might be almost invisible until closely sighted! Approximately 50,000 vessels of all types currently pass through the Strait of Gibraltar annually and the two ship traffic lanes are considered to be military chokepoints; at least one, and possibly two, Frei Otto-style tensioned fabric ship locks will be required to accommodate post-construction GSTB shipping traffic (Otto, 1967). Collapse of a blocking 1 m-high GSTB stretching across the 20 km breadth of this oceanic gateway won't necessarily be a disaster for the Mediterranean Sea Basin nations; more or less, it will be similar to a strong storm surge event with a constant 106,000 $m^3$/s incoming "tide" rippling rapidly eastwards towards Greece and Turkey to eventually inundate (by 1 m) 2.5 x $10^6$ $km^2$ of the Mediterranean Sea surface. A 1 m-high tsunami impacting the 1 m-high air-exposed eastern face of the GSTB will probably exert a momentary pressure of 10,000 $kg/m^2$, or about 200,000,000 kg overall; in other words, a new kind of flexible hydrostatic seawall installable at suitable locales—Gulf of Khambhat, India? (Gupta and Sharma, 1995)—will become effectively available to coastal macro-engineers (Cathcart, 2006; Barry 2006).

Other than collision events caused by errant ships and the cycling pressure changes of small intra-Basin tides, the most significant prospective structural integrity maintenance threat facing a Gibraltar Strait Textile Barrage are tsunami generated within the Basin or in the North Atlantic Ocean. For example, the tsunami caused by the 1 November 1755 Lisbon, Portugal seismic event caused a maximum 11 m wave run-up at Tarifa and 10 m at Tangier in Morocco; the Rock of Gibraltar's seaport was impacted with a maximum wave run-up of 2 m. The historically recorded past and the predicted future (Hills and Goda, 1998) related to tsunami impacts on the GSTB-served region means that tsunami momentum effects must be very carefully considered in great detail. Harold G. Loomis (2002) states that when a tsunami wave meets a semi-slack barrier the barrier "…takes some momentum from the wave and transfers it to…[the barrier if it can move, making it taut]…. A fair amount of momentum is bounced back to the ocean in various directions." If one GSTB is judged insufficient to withstand the forces applied (snap loading), then another paralleling GSTB can be installed since each is not as costly as a single Gibraltar Strait Bridge! Elongation of the GSTB's super-ropes under dynamic loading will dissipate some tsunami-deposited energy.

In the case of the GSTB, there is a quite interesting newly apprehended hydraulic effect (Rao, 2005) only partially comprehended by GSTB proponents: any North Atlantic Ocean tsunami overtopping the Gibraltar Strait Barrage will encounter a sudden 1 m hydraulic descent. **IS IT POSSIBLE A SUDDEN DESCENT OF FLOWING SEAWATER AT THE GSTB IMPEDENANCE IN A ONE-METER WATERFALL WILL EFFECTIVELY ATTENUATE A POTENTIALLY DEVASTATING TSUNAMI'S SUBSEQUENT RUN-UP ON 13,000 KM OF**

MEDITERRANEAN SEA BASIN COAST?  The acute observation by Rao (2005) of the remarkable effect of a sudden topographical drop on tsunami propagation hydrodynamics may be as instigative subsequently as the contemplations of Benjamin Franklin (1706-90) on the spreading of oil on a UK freshwater pond as well as the ocean's seawater (Tanford, 2004; Mertens, 2006).

When a tsunami wave impacts the GSTB, part of the wave's energy is transmitted through the GSTB, part is reflected from the GSTB and another part is absorbed in the various materials of which the GSTB is constructed.  Tsunami overtopping of the impermeable tensioned fabric dam with zero freeboard will cause a North Atlantic Ocean seawater hydraulic flow (from supercritical to subcritical) that induces GSTB vibrations (near-critical flow induced vibrations, vortex shedding and suction on the air-exposed "downstream" face of the dam-artwork).  In effect, the Mediterranean Sea close to the intact quivering/resonating GSTB will become a "stilling basin" that will dissipate the kinetic energy of the overtopping flow.  (A 1 m overtopping results in a temporary 106,000 $m^3$/s flow and a 2 m overtopping results in a 300,000 $m^3$/s waterfall.)  The optimum value of the drop height at the GSTB must be determined mathematically and by physical model testing since macro-engineers will wish to stabilize its geographical position and not cause any dangerous structural damage to the GSTB. Unfortunately, R&D reports on tensioned membranes vertically spanning water depths of limited extent are infrequent in the appropriate literature (Lo, 2000).  The total area of the water-retaining fabric drape that is planned to comprise the GSTB is about 200 $km^2$ but only approximately 20,000 $m^2$ of it will actually be fully exposed to the air and material-degrading sunshine on its eastern face while under a continuous imposed 1 m seawater head (hydraulic pressure) on its submerged western face.

## 5. IMPERVIOUS MEMBRANE ARCHITECTURE

Cables and membranes are the essential main components of Frei Otto's architecture proposals. Application of advanced technical textiles and super-ropes composed mainly of Kevlar—or, eventually, even carbon nanotubes—could permit safe emplacement and use of a pontoon bridge spanning the Strait of Gibraltar—in fact, a vehicle-carrying floating bridge imitating the span used by Xerxes in 480 BC to support his marching troops and their baggage train as they crossed the Hellespont!  Braided or stranded super-ropes could stabilize a pontoon bridge in a fixed geographical alignment for a long period, especially in a 1 m $s^{-1}$ eastward flowing surface current refilling the evaporating Mediterranean Sea, which has a yearly seawater deficit of about 0.5 m.

The ultimate hydrostatic head supported by a textile (woven or non-woven) is the measure of the resistance to the passage of seawater through the material; the standard applicable for determining the resistance to seawater penetration is the hydrostatic pressure test.  Several international organizations, as well as many national agencies generally accept the height of a seawater column given in metric units of distance as the applicable validation of a test method primarily intended for dense fabrics and films.  Waterproof and watertight are synonyms in this report.  In the past, the resistance to seawater penetration (in ship sails, ship hatch-covers etc.) has been technically achieved by coating woven textiles with various waterproofing materials; watertight textiles can now be achieved by dense weaving of strong fibers.  Multi-axial multi-ply textiles are bonded by a loop system, consisting of one or more yarn layers stretched in parallel; yarn layers can have different spatial orientations and different yarn densities.  The combination of multi-directional fiber layers has been found by scientific laboratory and commercial factory testing to be remarkably capable of distributing extra-ordinarily high strain forces; multi-axial multi-ply textile structures are dimensionally stable in any direction and exhibit isotropic distribution of stress forces with uniform strain behavior.  Kevlar 29, 49, and 149 since 1971 is a Dupont, USA, trade name for aromatic polyamides with a Tensile Strength of >3 Gpa, a Failure Strain of 3% and a Material Density of ~1.4 $g/cm^3$ is a good example.  All extremely strong materials able to perform as unitary form-active structures ought to be investigated for use in the proposed GSTB.

The characteristic strength of a structural textile or film material must have a low probability (~5%) of not being reached during the lifetime of the material's use in the GSTB and the characteristic load must not have more than a 5% probability of being exceeded during the design lifetime of the GSTB. Potentially, embedded fiber-optic electronics—detectors, reporters and automated alarm actuators—ought to be used to monitor in real-time the super-ropes as well as the draped barrier, which must fit tightly to Gibraltar Strait's sea bottom and sidewalls to successfully fulfill its macroproject functions, giving instant alerts to immediately responsible shore-based supervisors of all developing GSTB structural problems related to the GSTB's safe and efficient performance. Should the GSTB be separated from its two sidewalls and sea bottom anchorage, swept away by an over-whelming tsunami after full loss of its structural integrity, then possibly it might be partly retrievable/salvageable and, if so, its quick post-failure replacement would "re-initiate" the artificial Mediterranean Sea reduction (by natural evaporation) GSTB macroproject in a manner timely. A clever collapse design could even optimize recovery of a broken GSTB's components—in other words, the GSTB might be constructed with a design philosophy including the possibility of semi-controlled collapsibility, even pre-planned folding!

Let us now consider the very inexpensive-to-construct GSTB mathematically. The width of the Gibraltar Strait at the place designated previously (in Section 4) is 20,000 m, the maximum depth is 900 m, with an average depth of ~450 m. The GSTB will have a water surface difference of 1 m. If the top of the Gibraltar Strait Textile Barrier is partially supported by pontoons floating on the North Atlantic Ocean, the installation may be utilized as a vehicular highway between Spain and Morocco. Sea-going ships arriving and departing the Mediterranean Sea Basin will bypass the GSTB by using sturdy fabric ship-locks built at each terminus of the bridge-dam. Our purpose in Section 5 is to estimate the needed materials (film, supporting super-ropes) as well as the hydropower output potential. A simple sketch of the GSTB is provided in Fig. 1:

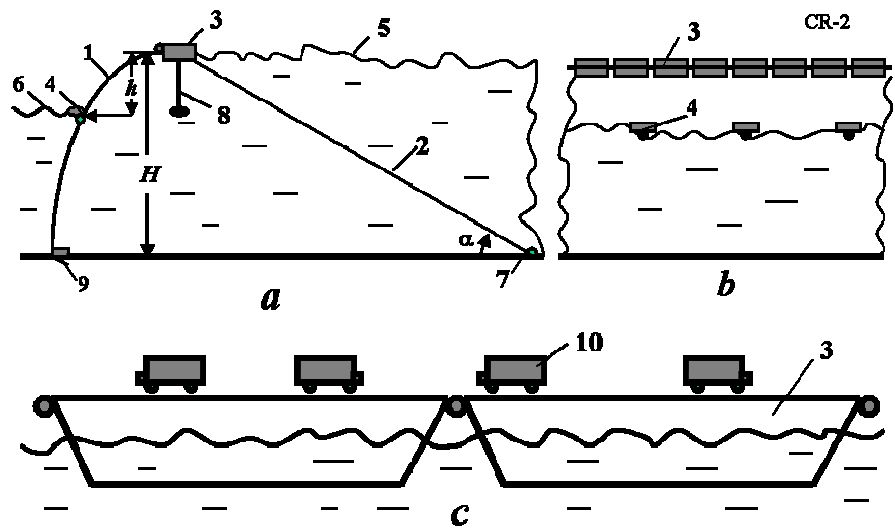

**Fig. 1.** Kevlar or other suitable film-like flexible material forming the seawater barrier and the installation's associated hydropower station. (a) side-view, (b) front-view, (c) pontoon highway bridge between Spain and Morocco. Notations: 1—flexible non-woven textile dam, 2—support cable, 3—pontoon, 4—hydroelectric turbine, 5—North Atlantic Ocean water level, 6—Mediterranean Sea water level, 7—anchor, super-rope spool, motor of support cable, 8—stabilizer, 9—stones, 10—tracks. Angle $\alpha = 30^0$.

## Computation

**1. Force** $P$ [N/m$^2$] for 1 m$^2$ of dam is

$$P = g\gamma h , \qquad (1)$$

where $g = 9.81$ m/s2 is the Earth gravity; $\gamma$ is water density, $\gamma = 1000$ kg/m3; $h$ is difference between top and lower levels of water surfaces, m (see computation in fig.2).

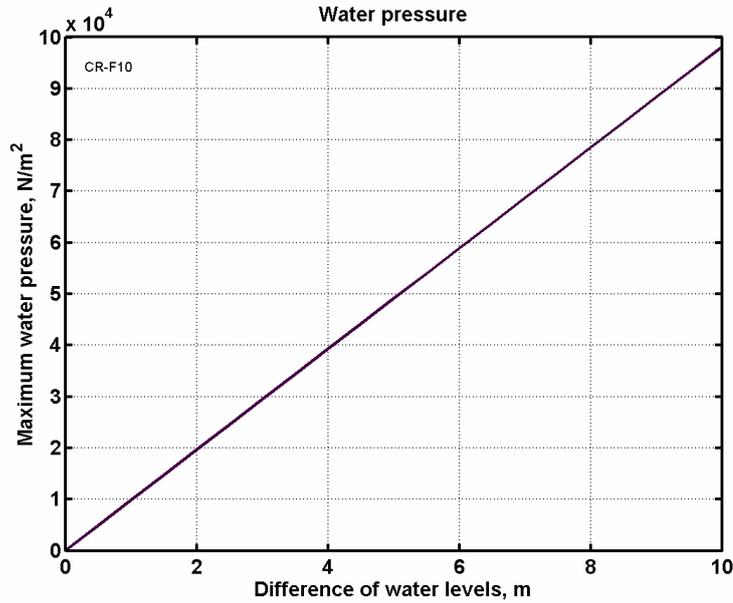

**Fig. 2**. Water pressure via difference of water levels

**2. Water power** $N$ [W] is

$$N = \eta g m h, \quad m = \gamma v S, \quad v = \sqrt{2gh}, \quad N = \eta g \gamma h S \sqrt{2gh}, \quad N/S \approx 43.453 \eta h^{1.5}, \; [\text{kW/m}^2] \qquad (2)$$

where $m$ is mass flow across 1 m width kg/m; $v$ is water speed, m/s; $S$ is turbine area, m$^2$; $\eta$ is coefficient efficiency of the water turbine, $N/S$ is specific power of water turbine, kW/m$^2$.
Computation is presented in fig. 3

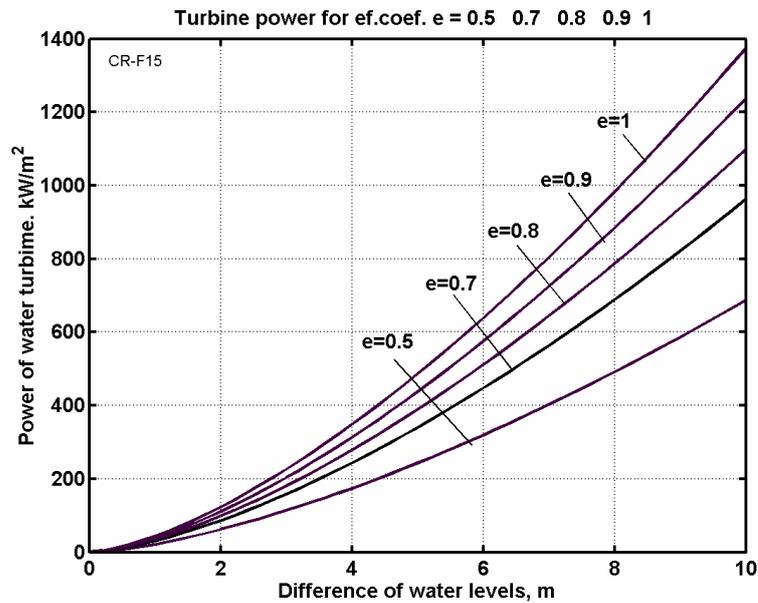

Fig. 3. Specific power of a water turbine via difference of water levels and turbine efficiency coefficient

**3. Film thickness** is

$$\delta = \frac{g\gamma h^2}{2\sigma},\qquad(3)$$

where σ is safety film tensile stress, N/m². Results of computation are in fig. 4.

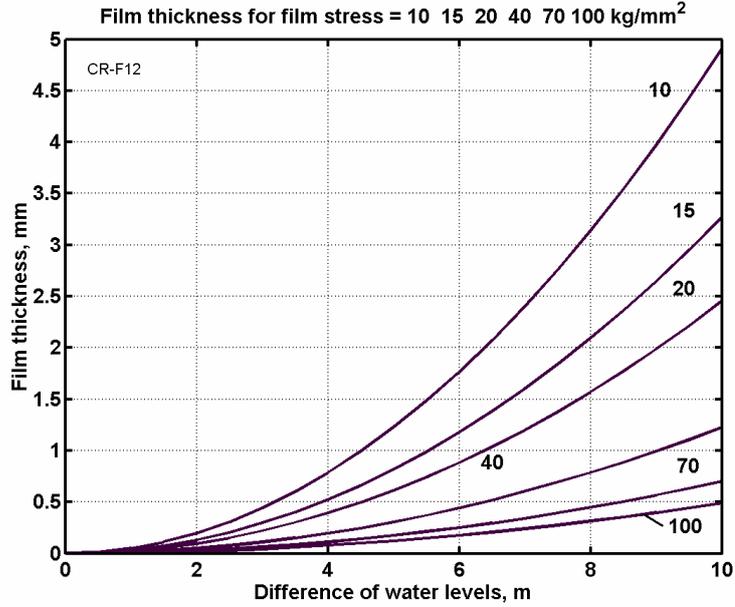

Fig. 4. Film thickness via difference of water levels safety film tensile stress.

**4. The film weight** of 1m width is

$$W_f = 1.2\delta\gamma H,\qquad(4)$$

Computation are in fig. 5. If our dam has long $L$ m, we must multiple this results by $L$.

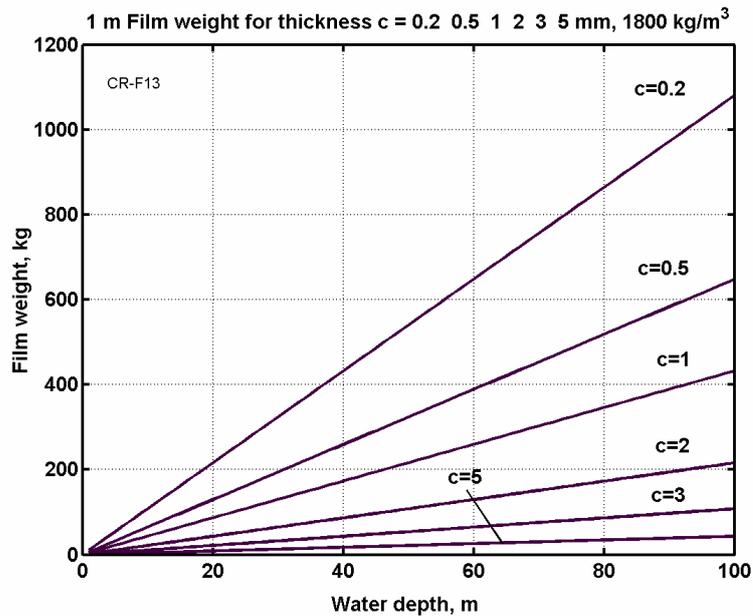

Fig. 5. Film weight via the deep of dam and film thickness $c$, density 1800 kg/m³.

**5**. **The diameter** *d* of the support cable is

$$T = \frac{Pl_2}{2}, \quad S = \frac{T}{\sigma}, \quad d = \sqrt{\frac{4S}{\pi}}, \tag{5}$$

where *T* is cable force, N; $l_2$ is distance between cable, m; *S* is cross-section area, m$^2$. Computation is presented in fig. 6. The total weight of support cable is

$$W_c \approx 2\gamma_c HSL/l_2, \quad W_a = \gamma_c SL, \tag{6}$$

where $\gamma_c$ is cable density, kg/m$^3$; *L* is length of dam, m; $W_a$ is additional (connection of banks) cable, m.

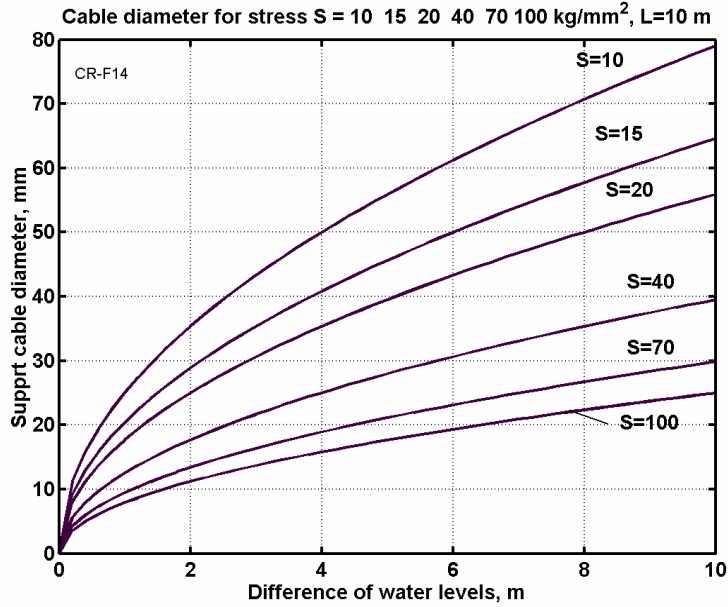

**Fig. 6**. Diameter of the support cable via difference of a water levels and the safety tensile stress.

Using the graphs above, we can estimate the relevant physical parameters of many interesting macroprojects.
Let us consider the GSTB emplaced between Spain and Morocco in the Strait of Gibraltar. The width of channel is 20000 m, the maximum deep is 900 m whilst the average depth is 450 m and the surface seawater level difference is 1m.
If the topmost part of the GSTB, on the North Atlantic Ocean side, is partially supported by pontoons, the installation may be used as a fixed link super-highway, possibly even as a railroad bridge. Transiting ocean-going ships can use fabric-gate sea-locks to cross the GSTB.
Our suggested Gibraltar strait Textile Barrier is very cheap to construct, especially when compared with a conventional concrete gravity dam installation as first proposed during 1928 by Herman Sorgel.

Our purpose is estimation of the needed film, supporting super-ropes, and energy development potential. A simple sketch of the GSTB is presented in fig. 1. The assumed difference of top and lower water levels (head) is 1 m. The estimation may be made using figs. 2 ÷ 6. However, we offer a clarifying computation that any reader can appreciate in this non-standard macro-engineering effort.
The force on 1 m$^2$ of film is $P = g\gamma h = 10 \times 1000 \times 1 = 10^4$ N/m$^2$, Eq.(1). The power 1 m of dam width for $\eta = 0.9$ is $N = \eta Pv \approx 4 \times 10^4$ W/m. For a dam length of 20,000 m the total power is $N_t = 800$ MW.

That is power of middle-of-the-Gibraltar Strait electric station. The head is small (1m only), that way the energy development potential is not as much.

Let us take the safety film and cable tensile stress $\sigma = 10$ kg/mm$^2$ $=10^8$ N/m$^2$. Using Eq. (3) we find the film thickness $\delta = 0.05$ mm, total film weight is $W_f = \delta \gamma HL = 0.05 \times 10^{-4} \times 1800 \times 450 \times 20000 = 810$ tons. Assume the support cables located $l_2 = 10$ m one from other. The cross-section area of cable is $S = Pl_2/\sigma = 10^4 \times 10/10^8 = 10^{-3}$ m, diameter is $d = 36$ mm. The total cable weight is 3240 tons plus 32 tons of the connecting 20,000 m cable. The total weight of our GSTB installation (without floating pontoons) is W = 810+3240+32 = 4182 tons. Floats can be also made from non-woven textile film and additional weight will be small ($\approx 5 \div 10\%$). We used conventional film, cables from cheapest available plastic. The artificial fiber has maximum tensile stress $\sigma = 500 \div 620$ kg/mm$^2$. Conventional strength factor has value 3 - 6. If we will use the artificial fiber with safety stress $\sigma = 100$ kg/mm$^2$, the total weight of installation decreases by 10 times (up to 418 tons). Than is millions times less (and much, much cheaper) then a conventional reinforced concrete gravity dam. The pontoons may be used as bridge for trains and/or motor vehicles. Conventional bypass channels may be used for surface shipping.

## 6. GSTB HYDROPOWER GENERATION OPPORTUNITY

Adjacent to the super-rope sidewall seals and sea bottom anchorages of our proposed Gibraltar Strait Barrage macroproject, the landscapes of Spain and Morocco offer nearly ideal conditions for the emplacement of Siphonic Hydropower facilities. Low-head hydroelectric power plants of a recently perfected type can manufacture electricity using gearless air turbines moved by seawater falling 1 m from the North Atlantic Ocean to the reduced/stabilized Mediterranean Sea. There can be an efficient recovery of economic energy in the form of electricity from low-head seawater. The monetary construction costs of these two power plants will be reasonable, even less than that of the GSTB; in total, this type of GSTB might be as little as 10% of the cost of a massive 1930s-style Herman Sorgel (1885-1952) concrete gravity dam—that is, about USA $10 billions instead of USA $100 billions! All the machinery housed in these power plants ought to be made secure from damaging environmental elements—possibly even small tsunami—and be readily examined and repaired by Spanish and Moroccan crews working on solid ground! Researched in the UK by Professor M.J. French since 1989 (French and Widden, 2001), Siphonic Hydropower has evidently reached a near-perfection status and early marketing stage in its rigorous macro-engineering R&D.
[GOTO: http://www.engineering.lancs.ac.uk/REGROUPS/LUREG/home/content.asp ]
The GSTB will present Siphonic hydropower developers with all the basic elements required for a potential hydroelectric development—a "limitless" world-ocean stream and an anthropogenic "drop" through which the North Atlantic Ocean's seawater can be used to convert the potential hydraulic energy into electrical energy. Where the available seawater head for the GSTB is only 1 m the losses in the Siphonic system (for aeration, due to pipe friction and due to upward drift of the air bubbles) would waste probably 0.4 m of the available head. Thus, the efficiency of the Siphonic system would be about 60%, without considering losses in the air turbines and electrical transmission losses. A Siphonic system of the size envisioned for the GSTB would justify the use of very efficient—that is, costly and precisely manufactured—air turbines, say at least 80%, so that overall efficiency could be approximately 50%. Thus, the electricity output could be ~500 MW. While not remarkably large, the power would be generated 24 hours/day, every day of the year regardless of weather conditions, forming a Mediterranean Sea Basin electrical generation and distribution system's reliable base-load. One megawatt is enough to power 1,000 European-style homes. Compressed air turbines energized by falling seawater will not affect the region's migratory birds, which is an advantage over wind farms where windmills are clustered at a site with persistent favorable winds (Lucas et al., 2004).

# 7. AFTERWORD

We propose a unique Ocean Art macroproject—the combination of the Gibraltar Strait Textile Barrage and Siphonic Hydropower —as an inexpensive total solution for the major known and forecast multiple environmental macro-problems of the Mediterranean Sea Basin. Proof and practical adaptation of this technology at the GSTB might foster its adoption elsewhere. We have also offered a new kind of textile-based hydrostatic seawall capable of successfully tolerating unpredictable incident 1 m-high tsunami waves. Joseph-Marie Jacquard (1752-1834) invented the automatic loom and his work with calculating machines has eventuated in today's supercomputers, according to James Essinger's ***Jacquard's Web*** (Oxford University Press, 2004, 302 pages). How fitting, then, that commercially available textiles/super-ropes and computers are precisely the two industrial tools most needed to successfully resolve our *gedanken* experiment!